\documentclass[aps,prc,superscriptaddress,twoside,twocolumn,showpacs]{revtex4}
\usepackage{amsmath,amssymb}
\usepackage{graphicx}
\usepackage{graphics}
\usepackage{epsfig}

\begin{document}
\title{The production of $\boldsymbol{K^+K^-}$ pairs in proton-proton collisions below the $\boldsymbol{\phi}$ meson threshold}
\author{Q.~J.~Ye}\email{qy4@phy.duke.edu}
\affiliation{Department of Physics and Triangle Universities Nuclear
Laboratory, Duke University, Durham, NC 27708, USA}
\affiliation{Institut f\"ur Kernphysik and J\"ulich Centre for Hadron
Physics, Forschungszentrum J\"ulich, D-52425 J\"ulich, Germany}
\author{M.~Hartmann}\email{m.hartmann@fz-juelich.de}
\affiliation{Institut f\"ur Kernphysik and J\"ulich Centre for Hadron
Physics, Forschungszentrum J\"ulich, D-52425 J\"ulich, Germany}
\author{D.~Chiladze}
\affiliation{Institut f\"ur Kernphysik and J\"ulich Centre for Hadron
Physics, Forschungszentrum J\"ulich, D-52425 J\"ulich, Germany}
\affiliation{High Energy Physics Institute, Tbilisi State University,
GE-0186
Tbilisi, Georgia}
\author{S.~Dymov}
\affiliation{Physikalisches Institut, Universit{\"a}t
Erlangen-N\"urnberg, D-91058 Erlangen, Germany}
\affiliation{Laboratory of Nuclear Problems, Joint Institute for
Nuclear Research, RU-141980 Dubna, Russia}
\author{A.~Dzyuba}
\affiliation{High Energy Physics Department, Petersburg Nuclear
Physics Institute, RU-188350 Gatchina, Russia}
\author{H.~Gao}
\affiliation{Department of Physics and Triangle Universities Nuclear
Laboratory, Duke University, Durham, NC 27708, USA}
\author{R.~Gebel}
\affiliation{Institut f\"ur Kernphysik and J\"ulich Centre for Hadron
Physics, Forschungszentrum J\"ulich, D-52425 J\"ulich, Germany}
\author{V.~Hejny}
\affiliation{Institut f\"ur Kernphysik and J\"ulich Centre for Hadron
Physics, Forschungszentrum J\"ulich, D-52425 J\"ulich, Germany}
\author{A.~Kacharava}
\affiliation{Institut f\"ur Kernphysik and J\"ulich Centre for Hadron
Physics, Forschungszentrum J\"ulich, D-52425 J\"ulich, Germany}
\author{B.~Lorentz}
\affiliation{Institut f\"ur Kernphysik and J\"ulich Centre for Hadron
Physics, Forschungszentrum J\"ulich, D-52425 J\"ulich, Germany}
\author{D.~Mchedlishvili}
\affiliation{Institut f\"ur Kernphysik and J\"ulich Centre for Hadron
Physics, Forschungszentrum J\"ulich, D-52425 J\"ulich, Germany}
\affiliation{High Energy Physics Institute, Tbilisi State University,
GE-0186
Tbilisi, Georgia}
\author{S.~Merzliakov}
\affiliation{Institut f\"ur Kernphysik and J\"ulich Centre for Hadron
Physics, Forschungszentrum J\"ulich, D-52425 J\"ulich, Germany}
\affiliation{Laboratory of Nuclear Problems, Joint Institute for
Nuclear Research, RU-141980 Dubna, Russia}
\author{M.~Mielke}
\affiliation{Institut f\"ur Kernphysik, Universit\"at
M\"unster, D-48149 M\"unster, Germany}
\author{S.~Mikirtytchiants}
\affiliation{Institut f\"ur Kernphysik and J\"ulich Centre for Hadron
Physics, Forschungszentrum J\"ulich, D-52425 J\"ulich, Germany}
\affiliation{High Energy Physics Department, Petersburg Nuclear
Physics Institute, RU-188350 Gatchina, Russia}
\author{H.~Ohm}
\affiliation{Institut f\"ur Kernphysik and J\"ulich Centre for Hadron
Physics, Forschungszentrum J\"ulich, D-52425 J\"ulich, Germany}
\author{M.~Papenbrock}
\affiliation{Institut f\"ur Kernphysik, Universit\"at
M\"unster, D-48149 M\"unster, Germany}
\author{A.~Polyanskiy}
\affiliation{Institut f\"ur Kernphysik and J\"ulich Centre for Hadron
Physics, Forschungszentrum J\"ulich, D-52425 J\"ulich, Germany}
\affiliation{Institute for Theoretical and Experimental Physics,
RU-117218 Moscow, Russia}
\author{V.~Serdyuk}
\affiliation{Institut f\"ur Kernphysik and J\"ulich Centre for Hadron
Physics, Forschungszentrum J\"ulich, D-52425 J\"ulich, Germany}
\affiliation{Laboratory of Nuclear Problems, Joint Institute for
Nuclear Research, RU-141980 Dubna, Russia}
\author{H.~J.~Stein}
\affiliation{Institut f\"ur Kernphysik and J\"ulich Centre for Hadron
Physics, Forschungszentrum J\"ulich, D-52425 J\"ulich, Germany}
\author{H.~Str\"oher}
\affiliation{Institut f\"ur Kernphysik and J\"ulich Centre for Hadron
Physics, Forschungszentrum J\"ulich, D-52425 J\"ulich, Germany}
\author{S.~Trusov}
\affiliation{Institut f\"ur Kern- und Hadronenphysik,
Helmholtz-Zentrum Dresden-Rossendorf, D-01314 Dresden, Germany}
\affiliation{Skobeltsyn Institute of Nuclear Physics, Lomonosov Moscow
State University, RU-119991 Moscow, Russia}
\author{Yu.~Valdau}
\affiliation{Institut f\"ur Kernphysik and J\"ulich Centre for Hadron
Physics, Forschungszentrum J\"ulich, D-52425 J\"ulich, Germany}
\affiliation{Helmholtz-Institut f\"ur Strahlen- und Kernphysik,
Universit\"at
Bonn, D-53115 Bonn, Germany}
\author{C.~Wilkin}
\affiliation{Physics and Astronomy Department, UCL, London WC1E 6BT,
United Kingdom}
\author{P.~W\"ustner}
\affiliation{Zentralinstitut f\"ur Elektronik,
Forschungszentrum J\"ulich, D-52425 J\"ulich, Germany}
\date{\today}
\begin{abstract}
The $pp\to ppK^+K^-$ reaction was measured below the $\phi$ threshold at a
beam energy of 2.568~GeV using the COSY-ANKE magnetic spectrometer. By
assuming that the four-body phase space is distorted only by the product of
two-body final state interactions, fits to a variety of one-dimensional
distributions permit the evaluation of differential and total cross sections.
The shapes of the distributions in the $Kp$ and $Kpp$ invariant masses are
reproduced only if the $K^-p$ interaction is even stronger than that found at
higher energy. The cusp effect in the $K^+K^-$ distribution at the
$K^0\bar{K}^0$ threshold is much more clear and some evidence is also found
for coupling between the $K^-p$ and $\bar{K}^0n$ channels. However, the
energy dependence of the total cross section cannot be reproduced by
considering only a simple product of such pair-wise final state interactions.
\end{abstract}
%
\maketitle
%
%
\section{Introduction}
\label{introduction}%

The original motivation for the study of kaon-pair production in the $pp\to
ppK^+K^-$ reaction near threshold was the investigation of the structure of
the scalar mesons $a_0(980)$ or $f_0(980)$~\cite{Oel91}. Such measurements
were initially performed by the COSY-11 collaboration at several different
excess energies below the $\phi$-meson production
threshold~\cite{Wol98,Que01,Win06}. However, their results showed that scalar
meson production cannot in fact be the dominant driving mechanism in kaon
pair production~\cite{Que01} and that the data can be explained without the
explicit inclusion of the $a_0/f_0$. Furthermore, they showed that the $K^-p$
and $K^-pp$ invariant mass spectra were strongly distorted, presumably by the
$K^-p$ final state interaction (FSI)~\cite{Win06}. This was most apparent in
the ratio of the differential cross sections in terms of the $K^-p$ and
$K^+p$ invariant masses.

The $pp\to ppK^+K^-$ reaction was also investigated with higher statistics
above the threshold for the production of the $\phi$ meson, mainly with the
aim of investigating the properties of that
meson~\cite{Bal01,Mae08,Dzy08,QYe12}. After removing the $\phi$ contribution
in the spectra, it was clear that the $K^-p$ and $K^-pp$ distributions in the
non-$\phi$ data were both strongly influenced by the $K^-p$
interaction~\cite{Mae08,QYe12}. It has been suggested that this is connected
with the production of the $\Lambda(1405)$ excited hyperon~\cite{Wil09},
which might be treated as a $\bar{K}N$ quasi-bound state with a width that
overlaps the $\bar{K}N$ threshold~\cite{Dal59}. This idea was put on a
quantitative footing by assuming that the $\Lambda(1405)$ was formed through
the decay $N^\star\to K^+\Lambda(1405)$~\cite{Xie10}. The strength and
details of the $\bar{K}N$ interaction are clearly important elements in the
interpretation of possible kaon nuclear systems, such as the deeply bound
$K^-pp$ states~\cite{Yam10}.

In addition to the $K^-p$ FSI, and one between the two protons, the data also
showed an enhancement at low $K^+K^-$ invariant masses with some possible
structure at the $K^0\bar{K}^0$ threshold~\cite{Mae08,Dzy08,QYe12}. Though
the effects are small, they might be influenced by the $a_0(980)$ or
$f_0(980)$ scalar mesons. However, the investigation of this region was
hampered by the need to separate the non-$\phi$ from the $\phi$ contribution
and the fact that the data were spread over a very wide range of $K^+K^-$
invariant masses. Measurements below the $\phi$ threshold can provide useful
information on these interesting FSI effects without suffering the distortion
of the $\phi$ meson. However, the limited statistics in the low-energy
COSY-11 data~\cite{Wol98,Que01,Win06} are insufficient for detailed studies.

Previous measurements of the $pp\to ppK^+K^-$ reaction were carried out at
the COSY-ANKE magnetic spectrometer at $\varepsilon = 51$, 67, and
108~MeV~\cite{Mae08,QYe12}, where the $\phi$ threshold is at
$\varepsilon=32.1$~MeV. Here the excess energy is defined as $\varepsilon =
\sqrt{s}-2(m_p+m_K)c^2$, where $\sqrt{s}$ is the total center-of-mass energy and
$m_p$ and $m_K$ are the particle masses in the final state. Because of the
limited acceptance of this spectrometer, an ansatz has to be made regarding
the distribution of events over the four-body phase space in order to convert
count rates into cross sections. This was done assuming that the distortions
were the products of those present in the two-particle subsystems. All the
ANKE non-$\phi$ data seemed to be consistent with an \emph{effective}
scattering length of $a_{K^-p}=(0+1.5i)$~fm with no obvious influence of an
energy dependence associated with an effective range term. The dominance of
the imaginary part is not unexpected because of the strong couplings to the
$\Sigma\pi$ and $\Lambda\pi$ channels but, due to the presence of two other
final-state particles, this parameter is not necessarily an intrinsic feature
of the isolated $K^-p$ system.

The ANKE measurements at three excess energies also showed some enhancement
at low $K^+K^-$ invariant masses but with at least a break of slope at the
$K^0\bar{K}^0$ threshold. A combined analysis of all the results in this
region~\cite{Dzy08,QYe12} shows that the data can be understood in terms of a
final state interaction involving both $K^+K^-$ elastic scattering plus a
contribution from the $K^+K^- \rightleftharpoons K^0\bar{K}^0$ charge
exchange. Although suggestive, the data are not sufficient to draw firm
conclusions.

In this paper we present much more precise $pp\to ppK^+K^-$
differential cross section data at a beam energy of $T_p =
2.568$~GeV ($\varepsilon=23.9$~MeV) obtained using the
COSY-ANKE spectrometer. With high statistics on the reaction
below the $\phi$-meson threshold, we could study the effects of
the final state interactions in the $K^-p$ and $K^+K^-$ systems
in greater details.

The paper is organized as follows. We first describe the experimental setup
and data analysis in Sec~\ref{Experiment}. Given that the procedures involved
are similar to those employed at higher energies~\cite{Mae08,QYe12}, this can
be quite brief. The fitting of the phenomenological parametrization to the
raw $pp\to ppK^+K^-$ data in order to make acceptance corrections is also
described here. The resulting differential cross sections and total cross
section for the $pp\to ppK^+K^-$ reaction are presented in Sec~\ref{results},
followed by our conclusions in Sec~\ref{conclusions}.

\section{Experiment and data analysis}
\label{Experiment}%

The measurement of the $pp\to ppK^+K^-$ reaction was performed at an internal
target station of the Cooler Synchrotron (COSY) of the Forschungszentrum
J\"{u}lich~\cite{Mai97}.  The ANKE spectrometer~\cite{Bar01,Har07}, which
consists of three dipole magnets, registers positively and negatively charged
ejectiles in the side detection systems, with the fast positively charged
particles being detected in the forward detector. Particle identification
relies on time-of-flight measurements~\cite{Har06,Har07,Mae08,Bus02} from
START and STOP counters, and momentum information obtained from the multiwire
proportional chambers.

\begin{figure}[h!]
\centering
\includegraphics[width=0.9\columnwidth,clip]{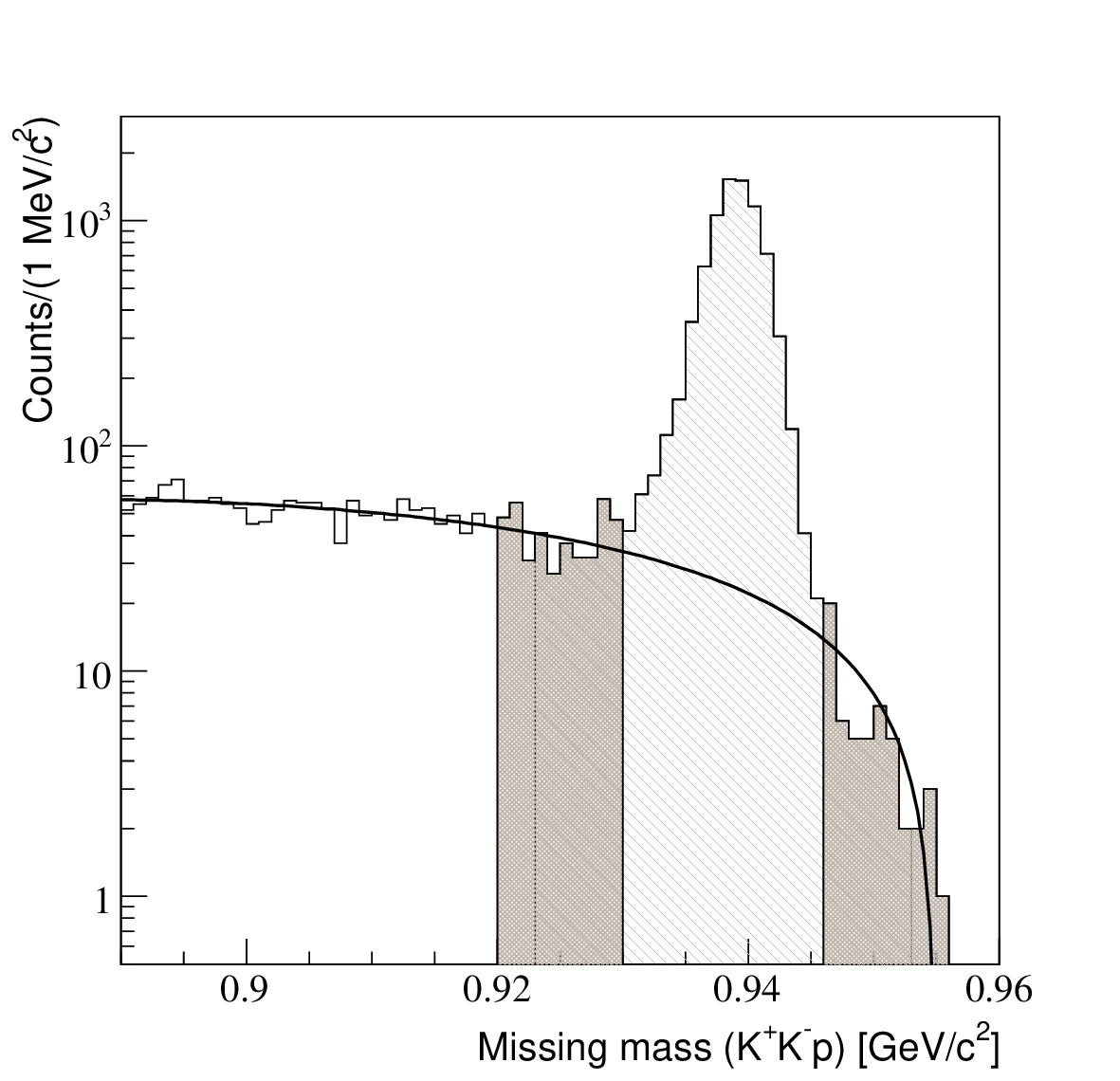}
\caption{The $pK^+K^-$ missing-mass distribution in the $pp\to pK^+K^-X$
reaction at $T_p = 2.568$~GeV. The hatched histogram shows the cuts
imposed for the selection of the non-detected proton. The solid line, which
is a second-order polynomial fit, was used to estimate the background
contribution under the proton peak.} \label{MM2.57}
\end{figure}

In close-to-threshold production experiments, the total cross section changes
very rapidly with small changes in the excess energy. The proton beam energy,
$T_{p} = 2.568$~GeV, was therefore determined very precisely through a
careful monitoring of the Schottky spectra~\cite{Ste08}. The resulting value
of the excess energy with respect to the $ppK^+K^-$ production threshold,
$\varepsilon=23.9$~MeV, is well below the nominal $\phi$ threshold. However
it should be noted that this is an average value, since the beam energy
decreases by up to 4.6~MeV through the course of a machine cycle due to the
interaction with the target. This effect was also investigated in the
simulation.

The experiment relied on a triple-coincidence, involving the observation of a
$K^+K^-$ pair in the side detectors and a fast proton in the forward
detector. The $pp\to ppK^+K^-$ reaction was then identified by requiring that
the missing mass of the $K^+K^-p$ system be consistent with that of a proton.
In the analysis, a $\pm3\sigma$ ($\sigma=2$~MeV/$c^2$) cut was applied on the
missing-mass distribution of the selected $K^+K^-p$ events, as shown in
Fig.~\ref{MM2.57}. The fraction of misidentified events inside the cut window
around the proton mass was estimated to be about 5\%, which was subtracted
from the peak using weighted data from the side bands, as parameterized by
the solid line. Any ambiguity in this procedure, which is less than 3\%, is
one source of systematic uncertainty.

\begin{figure}[h!]
\centering
\includegraphics[width=1.05\columnwidth,clip]{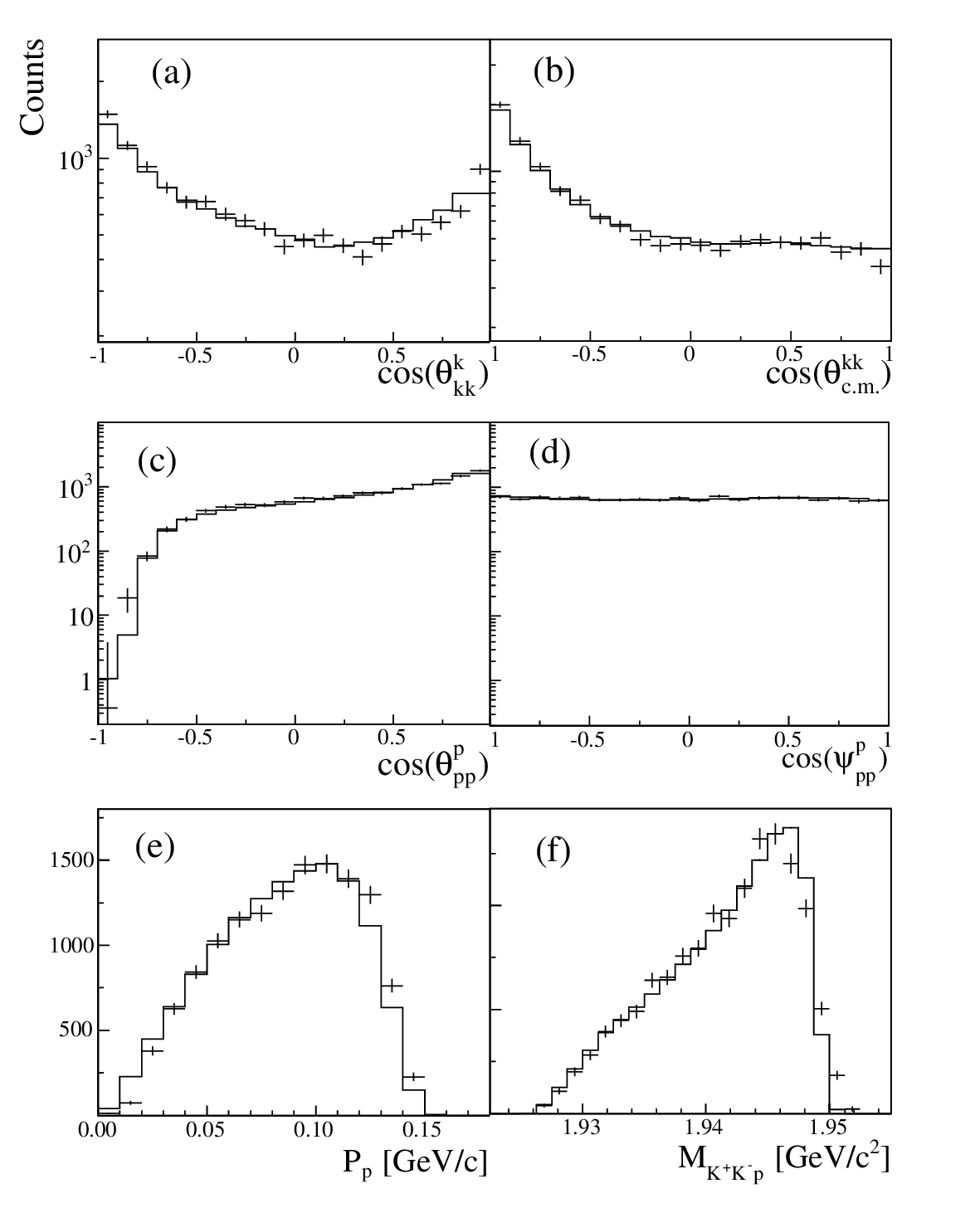}
\caption{Differential distributions of experimental (points) and simulated
(histograms) yields for kaon pair production in the $pp \to ppK^+K^-$
reaction at $\varepsilon=23.9$~MeV. Vertical bars represent the statistical
uncertainties and horizontal ones the bin widths. The individual panels are
(a) the cosine of the polar angle of the $K^+$ in the $K^+K^-$ reference
frame, (b) the polar angle of the kaon pairs in the overall c.m.\ frame, (c)
the polar angle of the emitted proton in the $pp$ reference frame relative to
the beam direction, (d) the polar angle of the proton in the $pp$ reference
frame relative to the direction of the kaon pair, (e) the proton momentum in
the $pp$ reference frame, and (f) the $K^+K^-p$ invariant mass.}
\label{Raw}
\end{figure}

After identifying clean $pp\to ppK^+K^-$ events in ANKE, acceptance
corrections must be performed in order to evaluate differential cross
sections. The simple ansatz used on data taken above the $\phi$ meson
production threshold tried to take into account the influence of final state
interactions in the various two-particle subsystems~\cite{Mae08,QYe12}. This
ansatz, which is also the basis for the current simulation, assumes that
the overall enhancement factor $F$ is the product of enhancements in the
$pp$, $K^+K^-$, and $K^-p$ systems:
\begin{equation}
\label{assume}
F = F_{pp}(q_{pp}) \times F_{Kp}(q_{Kp_{1}}) \times F_{Kp}(q_{Kp_{2}}) \times F_{KK}(q_{KK}),
\end{equation}
where $q_{pp}$, $q_{Kp_{1}}$, $q_{Kp_{2}}$, and $q_{KK}$ are the magnitudes of
the relative momenta in the $pp$, the two $K^-p$, and the $K^+K^-$ system,
respectively. It is believed that the $K^+p$ interaction might be weakly
repulsive and, if so, its neglect would be interpreted as extra attraction in
the $K^-p$ system. The FSI enhancement in the $K^-p$ case was calculated in
the scattering length approximation, $F_{Kp}(q)\approx 1/|1-iqa|^2$ and the
best fit to the higher energy data~\cite{Mae08,QYe12} was found with a purely
imaginary effective scattering length, $a_{K^-p}\approx 1.5i$~fm. The
proton-proton enhancement factor was derived from the Jost
function~\cite{Mae08,QYe12}. The enhancement factor in the $K^+K^-$ system
takes into account elastic $K^+K^-$ scattering plus the charge-exchange
$K^+K^- \rightleftharpoons K^0\bar{K}^0$~\cite{Dzy08}.

The seven degrees of freedom required to describe the unpolarized $ppK^+K^-$
final state were chosen to be four angles, the $K^+K^{-}$ and $K^+K^{-}p$
invariant masses, and the relative momentum of the protons in the $pp$
system~\cite{QYe12,Mae08}. Distributions in these seven variables were
generated inside the ANKE acceptance and compared with the experimental data,
some of which are shown in Fig.~\ref{Raw}. The best fit to the data was
achieved with $a_{K^-p}=(2.45\pm0.4)i$~fm, which is significantly larger in
magnitude than the starting value of $a_{K^-p}=1.5i$~fm. The uncertainty in
the real part is large and strongly correlated with the imaginary part. To
allow easy comparison with the analysis of the higher energy
data~\cite{Mae08,QYe12}, the effective scattering length was taken to be
purely imaginary.

The $\bar{K}K$ scattering lengths for isospin-one and zero were taken as in
our previous work~\cite{Dzy08} and the ratio of the $I=1$ and $I=0$
production amplitudes of $s$-wave $K\bar{K}$ pairs was parameterized as
$Ce^{i\phi_{c}}$. The best fit was obtained with $C= 0.54\pm0.03$ and
$\phi_{c} =-112^{\circ}\pm4^{\circ}$, which are consistent with our earlier
evaluation~\cite{QYe12,Dzy08} based on the above $\phi$ threshold data. The
resulting descriptions of the experimental data in Fig.~\ref{Raw} are very
good and certainly sufficient for evaluating the acceptance corrections.

The luminosity needed in the analysis was determined with an overall
systematic uncertainty of 9\% by measuring $pp$ elastic scattering in the
forward detector~\cite{Mae08}. This was checked by simultaneous studies of
the beam current and Schottky spectra~\cite{Ste08}, which could fix the
absolute luminosity with a systematic uncertainty of 6\%. Within these
uncertainties the two methods agreed but, in order to be coherent with our
previous work, the luminosity extracted from the $pp$ elastic scattering data
was used in the final analysis.

\section{Results}
\label{results}%

\begin{figure}[h!]
\centering
\includegraphics[width=0.9\columnwidth,clip]{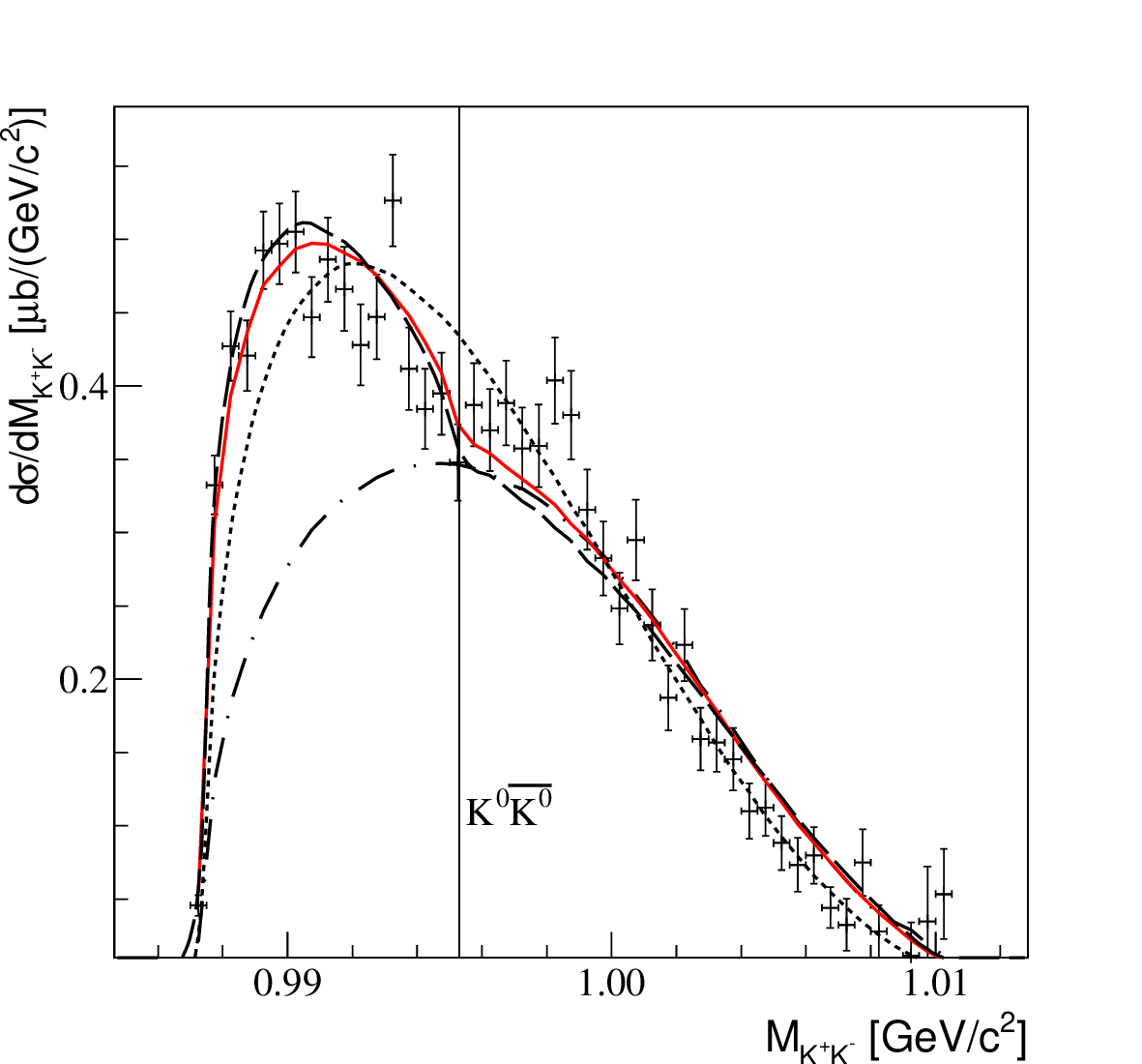}
\caption{(Color online) The $pp\to ppK^+K^-$ differential cross
section at $\varepsilon=23.9$~MeV as a function of the $K^+K^-$
invariant mass. The dotted curve shows the four-body phase
space simulation whereas the inclusion of the final state
interactions through Eq.~\eqref{assume} gives the dashed curve
for $a_{K^-p}=1.5i$~fm and the red solid curve
$a_{K^-p}=2.45i$~fm. The dot-dashed curve was obtained by
considering only the $pp$ and $K^-p$ final state interactions
with $a_{K^-p}=2.45i$~fm.} \label{IM}
\end{figure}

The differential cross section for the $pp\to ppK^+K^-$ reaction at an excess
energy $\varepsilon=23.9$~MeV is shown in Fig.~\ref{IM} as a function of the
$K^+K^-$ invariant mass. Also shown are simulations based on a four-body
phase space and this distorted by the final state interactions in the
$K^+K^-$, $pp$, and $K^-p$ systems within the product ansatz of
Eq.~\eqref{assume}. This was done separately with effective scattering
lengths of $a_{K^-p}=1.5i$~fm and $a_{K^-p}=2.45i$~fm.

\begin{figure}[h!]
\centering
\includegraphics[width=0.9\columnwidth,clip]{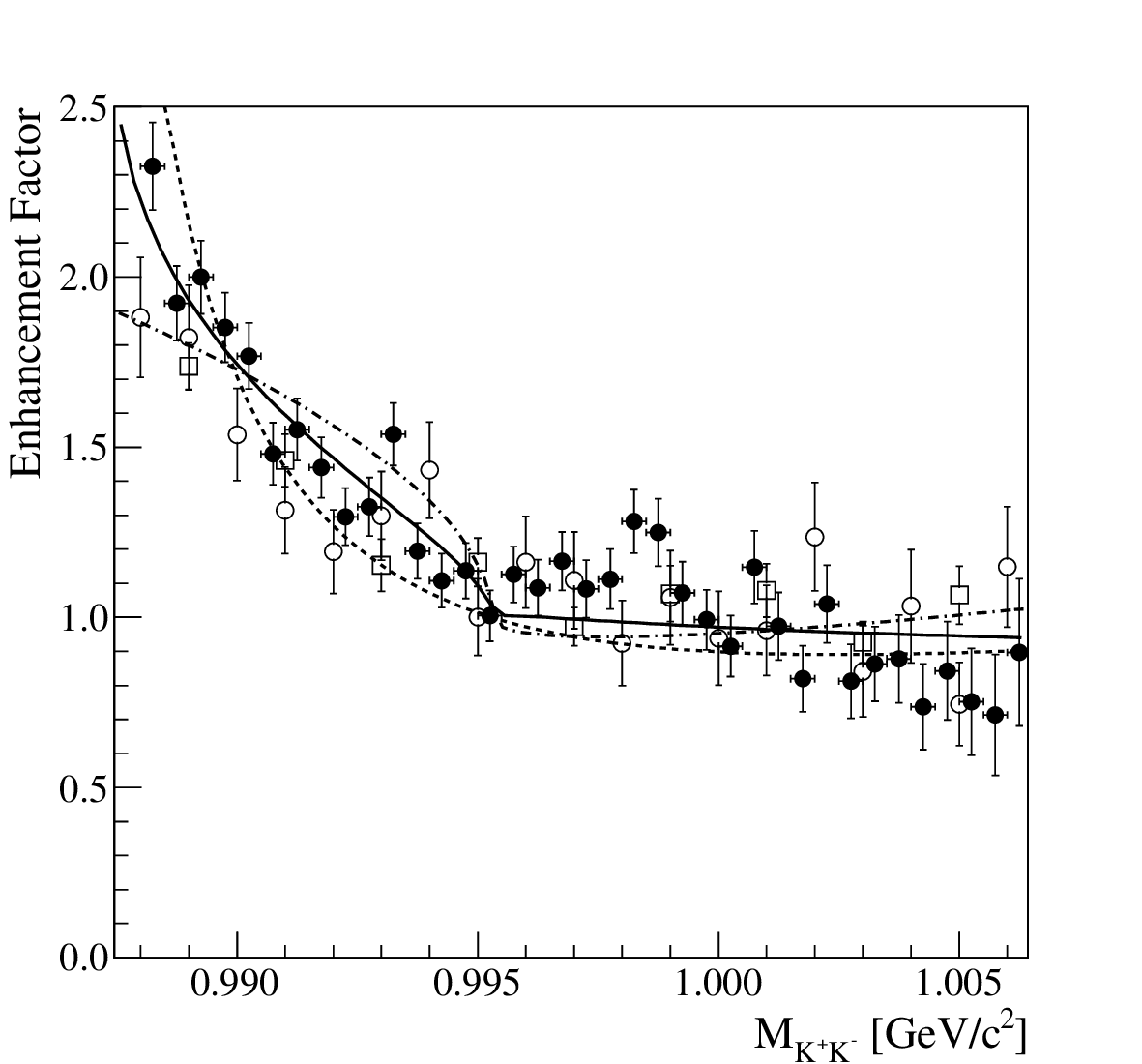}
\caption{Ratio of the measured $pp\to ppK^+K^-$ differential
cross section at $\varepsilon=23.9$~MeV as a function of the
$K^+K^-$ invariant mass to a simulation that includes only
$K^-p$ and $pp$ final state interactions (shown by the
dot-dashed curve in Fig.~\ref{IM}). In addition to the current
data (solid circles), weighted averages of previous
measurements (open squares and circles) are also presented. The
solid curve represents the best fit in a model that includes
elastic $K^+K^-$ FSI and $K^0\bar{K}^0 \rightleftharpoons
K^+K^-$ charge-exchange~\cite{Dzy08}. The best fits neglecting
charge exchange and including only this effect are shown by the
dashed and dot-dashed curves, respectively.} \label{Enhance}
\end{figure}

The most striking features in the data are the strength near the $K^+K^-$
threshold and the dip at $M_{K^+K^-} \approx 0.995$~GeV/$c^2$, which
corresponds precisely to the $K^0\bar{K^0}$ production
threshold~\cite{Dzy08}. This is compelling evidence for a cusp effect coming
from the $K^0\bar{K}^0 \rightleftharpoons K^+K^-$ transitions. To investigate
this phenomenon in greater detail, the $K^+K^-$ invariant mass distribution
was divided by a simulation where only the final state interactions in the
$pp$ and $K^-p$, with $a_{K^-p}=2.45i$~fm, were considered. The best fit to
the data shown in Fig.~\ref{Enhance} is achieved with a contribution from the
isospin-zero channel that is about three times stronger than the isospin-one.
This finding is consistent with our earlier result~\cite{Dzy08}. The
deviations apparent in Figs.~\ref{IM} and \ref{Enhance} at high $K^+K^-$
invariant masses might be connected with the approximations made in our
coupled-channel model~\cite{Dzy08}.

Previous analyses of the $pp\to ppK^+K^-$ reaction at different
excess energies~\cite{QYe12,Mae08,Win06,Sil09} have all shown a
strong preference for low values of the $K^-p$ and $K^-pp$
invariant masses, $M_{K^-p}$ and $M_{K^-pp}$. To study this
further, we have evaluated differential cross sections as
functions of these invariant masses and also the ratios:
\begin{eqnarray}
\label{IMratio}
&&  R_{Kp} = \frac{d\sigma/dM_{K^-p}}{d\sigma/dM_{K^+p}}, \nonumber\\
&&  R_{Kpp} = \frac{d\sigma/dM_{K^-pp}}{d\sigma/dM_{K^+pp}}\cdot
\end{eqnarray}

\begin{figure}[h!]
\centering
\includegraphics[width=0.9\columnwidth,clip]{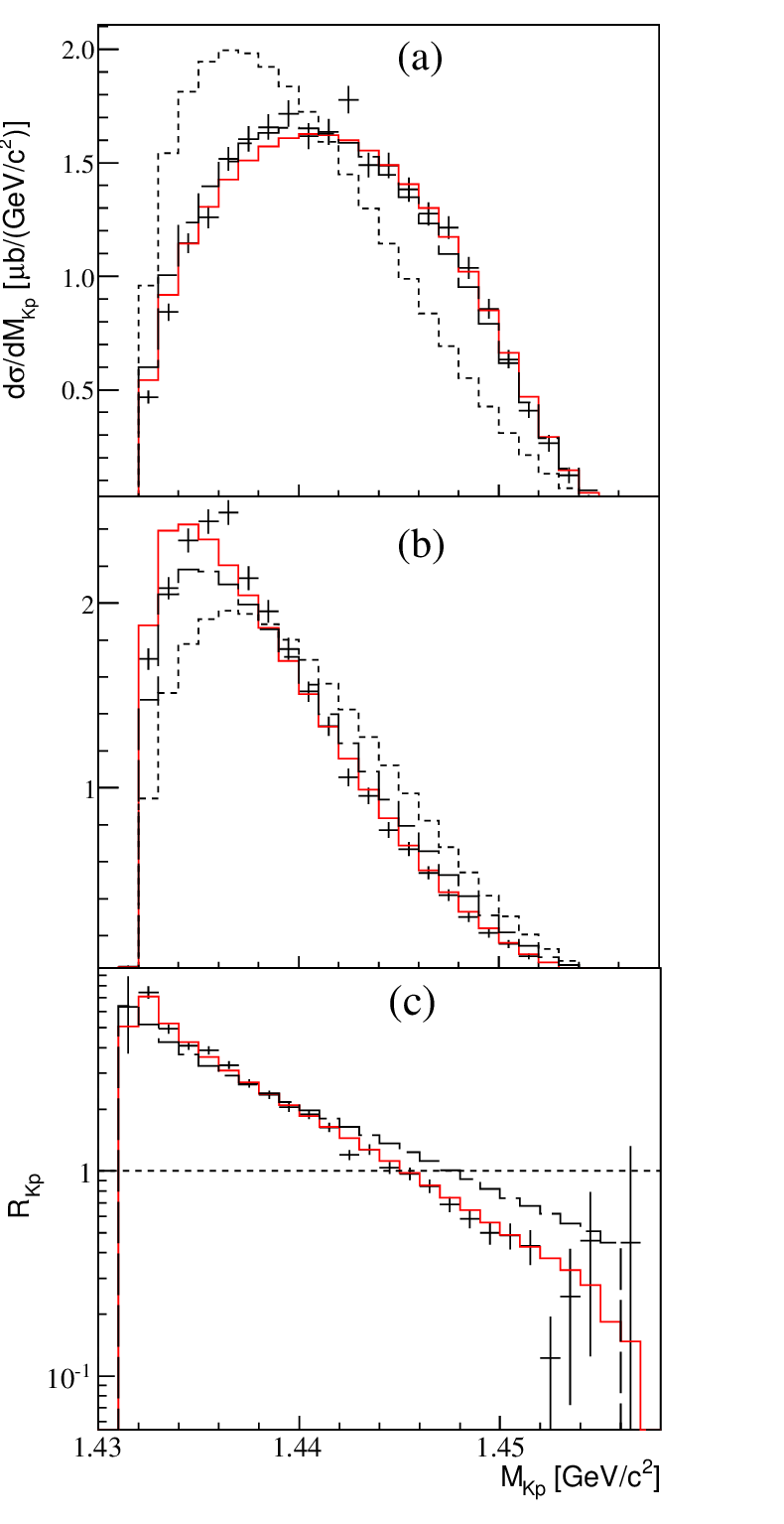}
\caption{(Color online) Differential cross sections for the
$pp\to ppK^+K^-$ reaction as functions of the invariant masses
of $K^+p$ (upper panel) and $K^-p$ (middle panel), and their
ratio $R_{Kp}$ (lower panel). The red solid and dashed black
histograms represent estimations based on Eq.~\eqref{assume}
that take into account $K^-p$, $pp$ and $K^+K^-$ final state
interactions with $a_{K^-p}=2.45i$~fm and $a_{K^-p}=1.5i$~fm,
respectively. The four-body phase-space simulations are shown
by the dotted histograms.} \label{Ratio_pK}
\end{figure}

\begin{figure}[h!]
\centering
\includegraphics[width=0.9\columnwidth,clip]{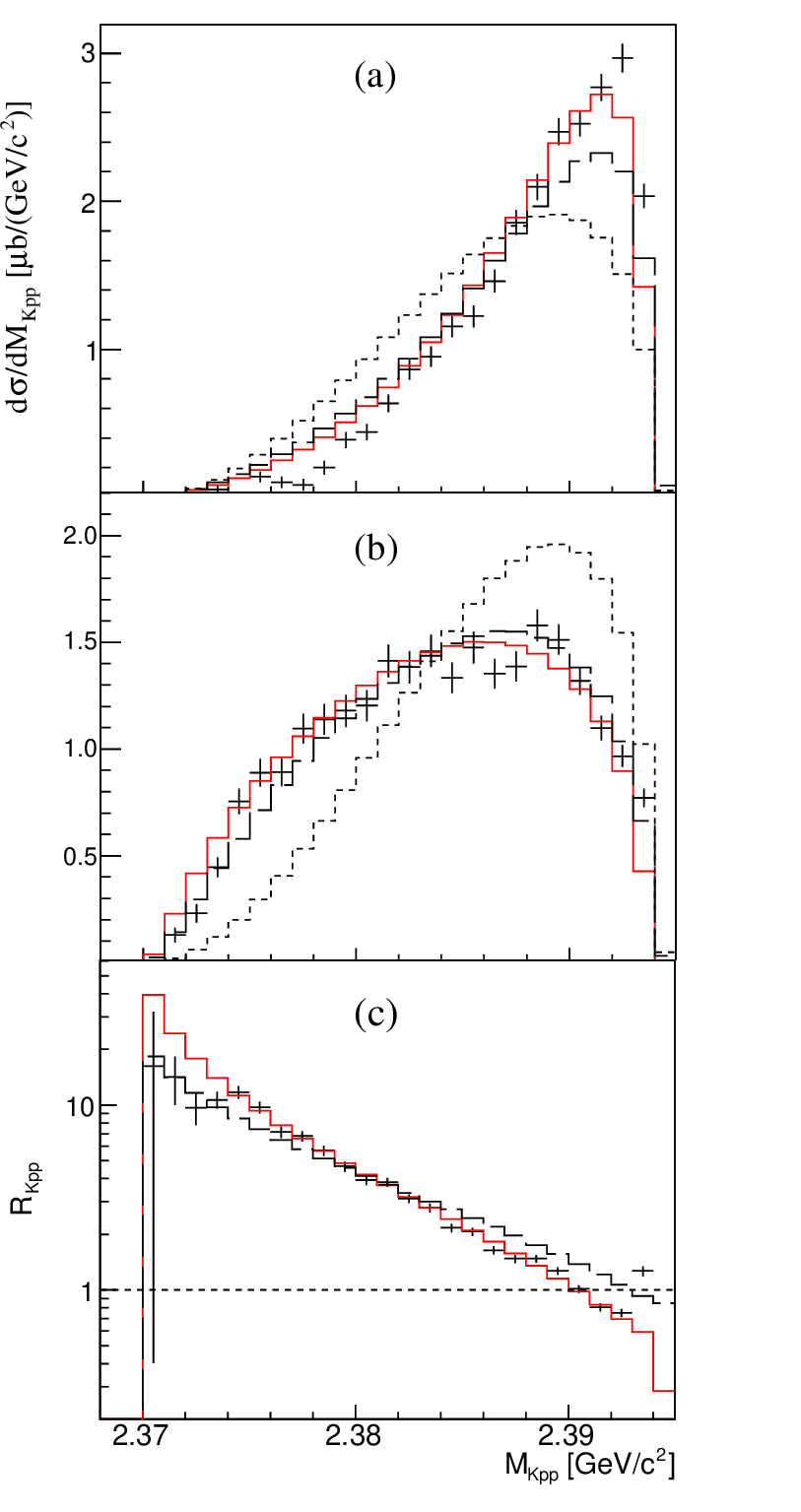}
\caption{(Color online) Differential cross sections for the $pp\to ppK^+K^-$
reaction with respect to the invariant masses of $K^+pp$ (upper panel) and
$K^-pp$ (middle panel), and their ratio $R_{Kpp}$ (lower panel). The
conventions for the theoretical estimates are as in Fig.~\ref{Ratio_pK}.}
\label{Ratio_ppK}
\end{figure}

The corresponding experimental data and simulations are shown
in Figs.~\ref{Ratio_pK} and \ref{Ratio_ppK}. Both $R_{Kp}$ and
$R_{Kpp}$ display the very strong preferences for lower
invariant masses seen in the earlier data. The low mass
enhancements in Figs.~\ref{Ratio_pK}c and \ref{Ratio_ppK}c
clearly indicate once again that the $pp\to ppK^+K^-$ reaction
cannot be dominated by the undistorted production of a single
scalar resonance $a_0$ or $f_0$. Within a four-body phase space
simulation both ratios should be constant and equal to one and
such a simulation also fails to describe the $M_{Kp}$ and
$M_{Kpp}$ distributions. Whereas the inclusion of a $K^-p$ FSI
with an effective scattering length $a_{K^-p}=1.5i$~fm improves
the situation, it overestimates the data in the high invariant
mass regions for both $R_{Kp}$ and $R_{Kpp}$. With the larger
effective scattering length $a_{K^-p}=2.45i$~fm, these ratios,
as well as the individual differential cross sections, can be
well reproduced. Within the product ansatz of
Eq.~\eqref{assume} the $K^-p$ final state interaction
effectively becomes stronger at lower excess energies. This
illustrates the limitations of this simple ansatz to the
complex four-body dynamics.

Although the $K^-p$ elastic final state interaction describes well the vast
bulk of the data shown in Figs.~\ref{Ratio_pK} and \ref{Ratio_ppK}, it is
interesting to note that there seems to be a small but significant deviation
between the $K^-p$ data and simulation in Fig.~\ref{Ratio_pK}b at low
invariant masses. Since the $\bar{K}^0n$ threshold is at 1.437~GeV/$c^2$,
this suggests that the data in this region might also be influenced by $K^-p
\rightleftharpoons \bar{K}^0n$ channel coupling.

\begin{figure}[h!]
\centering
\includegraphics[width=0.8\columnwidth,clip]{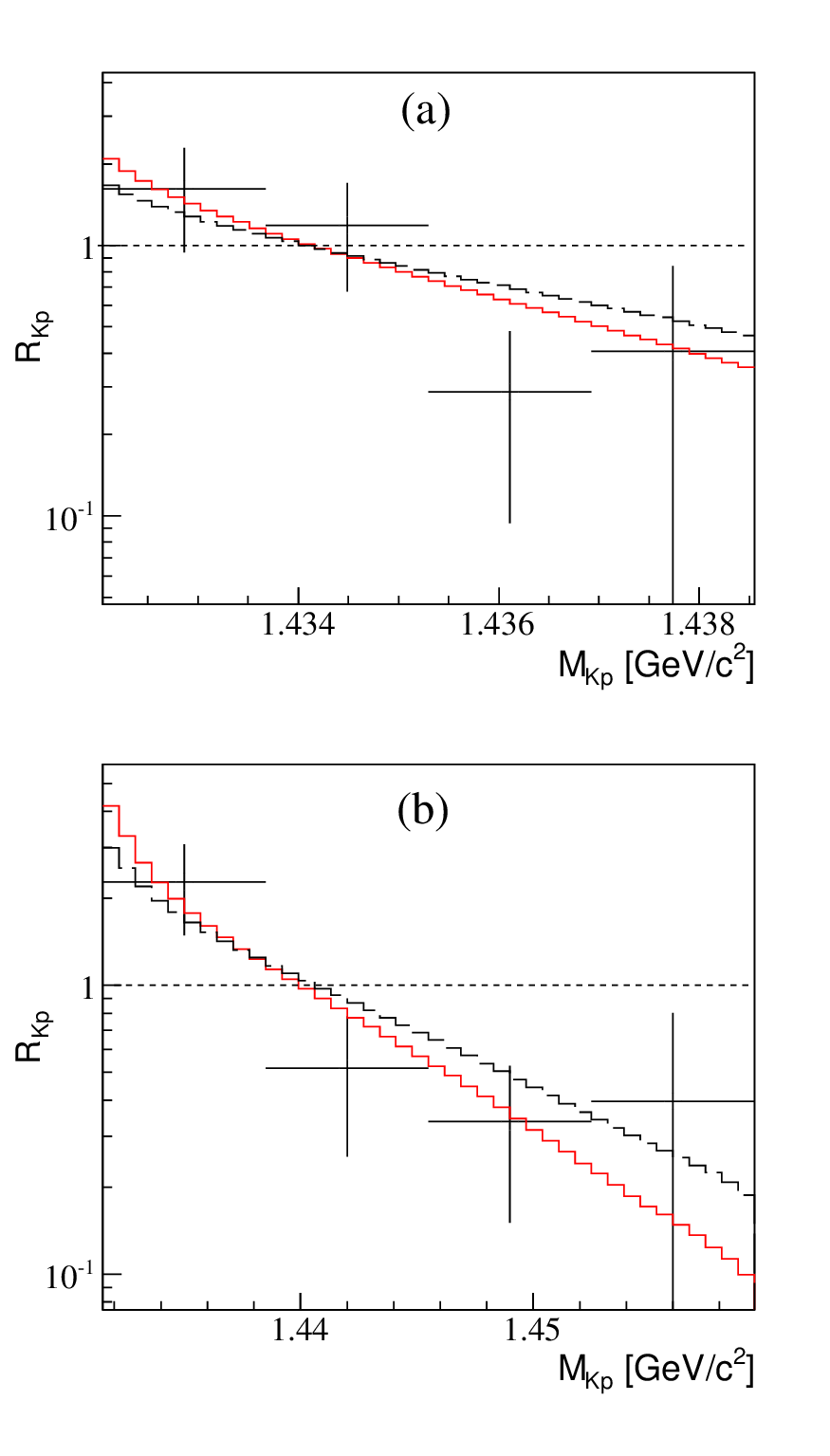}
\caption{(Color online)  The ratio $R_{Kp}$ for the $pp\to
ppK^+K^-$ reaction measured by COSY-11 at (a) $\varepsilon =
10$~MeV and (b) 28~MeV~\cite{Sil09}. The dotted histograms
represent the four-body phase-space simulations, whereas the
red solid and dashed ones represent the theoretical
calculations taking into account $K^-p$, $pp$ and $K^+K^-$
final state interactions with $a_{K^-p}=2.45i$~fm and
$1.5i$~fm, respectively.} \label{Ratio_10_28}
\end{figure}

Due to the low statistics, the COSY-11 data at 10 and
28~MeV~\cite{Win06,Sil09} cannot distinguish between predictions based on
effective scattering lengths of $a_{K^-p}=1.5i$~fm and $a_{K^-p}=2.45i$~fm.
This illustrated for the $R_{Kp}$ ratio in Fig.~\ref{Ratio_10_28} but this
lack of sensitivity is equally true for $R_{Kpp}$.

The $pp\to ppK^+K^-$ differential cross section, shown in
Fig.~\ref{IM} as a function of the $K^+K^-$ invariant mass, was
used to determine the value of the total cross section,
$\sigma=6.66\pm0.08\pm0.67$~nb, where the first error is
statistical and the second systematic. The systematic effects
considered here arise from the background subtraction,
acceptance correction, tracking efficiency correction, and
luminosity determination.

\begin{figure}[h!]
\centering
\includegraphics[width=0.9\columnwidth,clip]{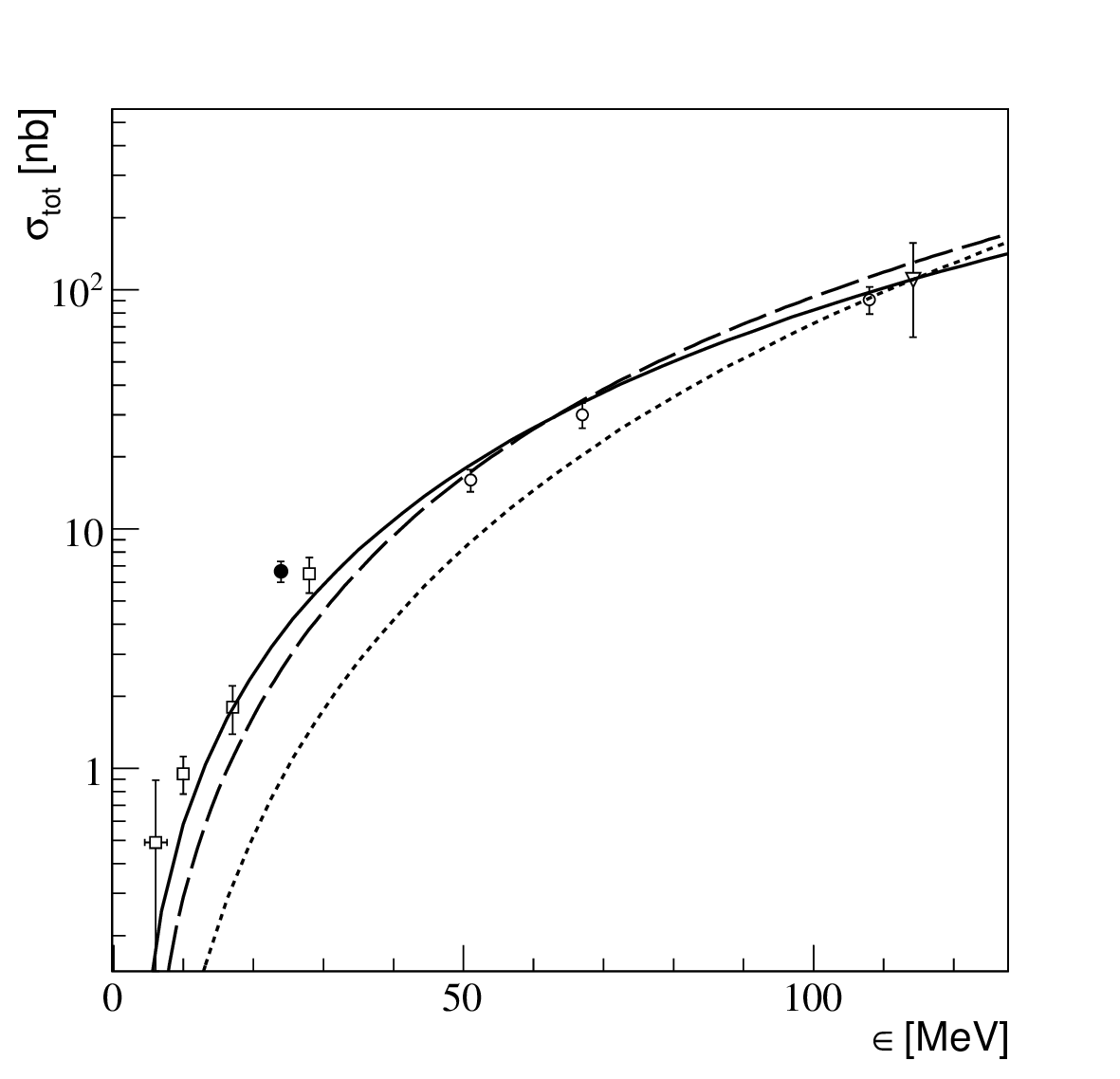}
\caption{Total cross section for the $pp\to ppK^+K^-$ reaction
as a function of excess energy $\varepsilon$. The present
result (closed circle) is shown together with earlier
experimental data taken from DISTO (triangle), ANKE (circles),
COSY-11 (squares). The dotted line shows the four-body phase
space simulation, whereas the solid line represent the
simulations with $a_{K^-p}=1.5i$~fm. The predictions of a
one-boson exchange model are represented by the dashed
line~\cite{Sib97}. } \label{txs}
\end{figure}

The total cross section result is plotted in Fig.~\ref{txs}
along with previous measurements from DISTO~\cite{Bal01},
COSY-11~\cite{Win06,Sil09,Wol98,Que01}, and
ANKE~\cite{Mae08,QYe12}. The new point seems high compared to
the COSY-11 result at $\varepsilon = 28$~MeV, though one has to
take into account the limited statistics of these data. This
value had already been increased by 50\% compared to that
originally published~\cite{Win06}. This was achieved through a
re-analysis of the data that included a modified $pp$ and a
$K^-p$ final state interaction with
$a_{K^-p}=1.5i$~fm~\cite{Sil09}. For the lower excess energy of
$\varepsilon = 10$~MeV, where the acceptance of the COSY-11
apparatus is higher, the re-analysis increased the cross
section by only 20\%. Both cross sections would be reduced
slightly if $a_{K^-p}$ were increased to $2.45i$~fm but the
changes would be less than the statistical errors~\cite{Sil13}.
The COSY-11 acceptance is very sensitive to the form assumed
for the $pp$ FSI but much less so for that of the
$K^-p$~\cite{Sil13}.

It is clear from Fig.~\ref{txs} that the four-body phase space
cannot reproduce the energy dependence of the total cross
section. With the inclusion of the $pp$, $K^+K^-$, and $K^-p$
FSI, with an effective scattering length of $a_{K^-p}=1.5i$~fm,
the data above the $\phi$ threshold can be described well but
those at lower energy are significantly underestimated. An
increase in the value of $a_{K^-p}$ might help in this region
but the coincidence of strong effects in different two- or even
three-body channels must also bring the factorization
assumption of Eq.~\eqref{assume} into question. The dashed
line, which represents a calculation within a one-boson
exchange model~\cite{Sib97}, also underestimates the
near-threshold data. This model includes energy-dependent input
derived from fits to the $K^{\pm}p\to K^{\pm}p$ total cross
sections, though it does not include the $pp$ final state
interaction.

\section{Discussion and Conclusions}
\label{conclusions}

The production of $K^+K^-$  pairs has been measured in the $pp\to ppK^+K^-$
reaction channel at an excess energy of $\varepsilon=23.9$~MeV. Even taking
its 4.3~MeV/$c^2$ width into account, this is well below the central
$\phi$-meson threshold at 32.1~MeV. The reaction was identified in ANKE
through a triple coincidence of a $K^+K^-$ pair and a forward-going proton,
with an additional cut on the $K^+K^-p$ missing-mass spectrum. The high
statistics and low excess energy allow us to produce a detailed $K^+K^-$
invariant mass distribution below the $\phi$ threshold.

The distortion of both the $K^-p$ and $K^-pp$ spectra, which are even
stronger than in our higher energy data, can be explained quantitatively
within the product ansatz of Eq.~\eqref{assume} with an effective $K^-p$
scattering length $a_{K^-p}\approx 2.45i$~fm. This is to be compared with the
$1.5i$~fm obtained from the analysis of data measured above the $\phi$
production threshold. A full treatment of the dynamics of the four-body
$ppK^+K^-$ channel is currently impractical. As a consequence, an energy
dependence of $a_{K^-p}$ is possible because this is merely an
effective parameter within a very simplistic description of the four-body
final state interaction. The strong $K^-p$ final state interaction may be
connected with the $\Lambda(1405)$ in the production process
and it has been suggested~\cite{Xie10} that the production of non-$\phi$ kaon
pairs proceeds mainly through the associated production $pp\to
K^+p\Lambda(1405)$. This would also lead to deviations from the simple
product ansatz for the final state interactions, not least because an
attraction between the $\Lambda(1405)$ and the proton would involve three
final particles.

Our results show a very strong preference for low $K^-pp$ masses and this
effect seems to be even more marked than in the higher energy
data~\cite{Mae08,QYe12}. Although this might be connected with the ideas of a
$K^-pp$ deeply bound state~\cite{Agn05,Yam07,She07,Yam10}, it must be
stressed that our data were measured far above threshold. They should not
therefore be taken as necessarily implying that the $K^-$ will bind with two
protons.

There is strong evidence for a cusp effect arising from the
$K^0\bar{K}^0 \rightleftharpoons K^+K^-$ transitions. Our
analysis within a coupled-channel description suggests that,
with the values of the $K\bar{K}$ scattering lengths used, the
production of isospin-zero $K\bar{K}$ pairs dominates. Though
this is consistent with results extracted from data taken above
the $\phi$ threshold~\cite{Dzy08,QYe12}, there is clearly room
for some refinement in the model. On the other hand, the
structure of the $K^-p$ invariant mass spectrum of
Fig.~\ref{Ratio_pK} in the 1437~MeV/$c^2$ region suggests that
there might be important coupling also between the $K^-p$ and
$\bar{K}^0n$ systems.

It is evident that the interactions in the four-body $ppK^+K^-$
final state are extremely complex. Nevertheless, the energy
dependence of the total cross section can be well described
above the $\phi$ threshold by introducing the effects of the
$pp$, $K^+K^-$ and $K^-p$ final state interaction with an
effective scattering length of $a_{K^-p}=1.5i$~fm. This would,
however, have to be increased to have any hope of fitting the
lower energy data. Further theoretical work is required to
clarify the reaction mechanisms.

%
\begin{acknowledgments}
We would like to thank the COSY machine crew for providing the
good conditions that were necessary for this work. We are also
grateful for the continued assistance of other members of the
ANKE Collaboration. Discussions with P.~Moskal and M.~Silarski
were very helpful. This work was supported in part by the US
Department of Energy under Contract No.~DE-FG02-03ER41231 and
COSY FFE.
\end{acknowledgments}
%
%

\end{document}